\documentclass[5p]{elsarticle}

\usepackage{algorithm}
\usepackage{algpseudocode}
\usepackage{amsmath,amsfonts}
\usepackage{cite}
\usepackage{graphicx}
\usepackage{lineno, hyperref}
\usepackage{textcomp}
\usepackage{subcaption}

\journal{Astronomy and Computing}

\biboptions{authoryear}

\begin{document}
\begin{frontmatter}

\title{PySAP: Python Sparse Data Analysis Package for Multidisciplinary Image Processing}

\author[cosmostat,neurospin]{S. Farrens}
\author[neurospin]{A. Grigis}
\author[neurospin]{L. El Gueddari}
\author[cosmostat,neurospin]{Z. Ramzi}
\author[neurospin]{Chaithya G. R.}
\author[epita]{S. Starck}
\author[ensta]{B. Sarthou}
\author[neurospin]{H. Cherkaoui}
\author[neurospin]{P. Ciuciu}
\author[cosmostat]{J.-L. Starck}

\address[cosmostat]{AIM, CEA, CNRS, Universit\'e Paris-Saclay, Universit\'e Paris Diderot, Sorbonne Paris Cit\'e, F-91191 Gif-sur-Yvette, France}
\address[neurospin]{CEA/DRF/Joliot NeuroSpin, Universit\'e Paris-Saclay, F-91191 Gif-sur-Yvette, France}
\address[epita]{EPITA, 14-16 rue Voltaire, 94270 Le Kremlin-Bic\^etre, France}
\address[ensta]{ENSTA ParisTech, 828 Boulevard des Mar\'echaux, 91120 Palaiseau, France}

\begin{abstract}
We present the open-source image processing software package PySAP (Python Sparse data Analysis Package) developed for the COmpressed Sensing for Magnetic resonance Imaging and Cosmology (COSMIC) project. This package provides a set of flexible tools that can be applied to a variety of compressed sensing and image reconstruction problems in various research domains. In particular, PySAP offers fast wavelet transforms and a range of integrated optimisation algorithms. In this paper we present the features available in PySAP and provide practical demonstrations on astrophysical and magnetic resonance imaging data.\end{abstract}


\end{frontmatter}


\section{Introduction}
\label{sec:intro}

The ability to obtain high quality data in a short amount of time or indeed to recover high resolution images from undersampled blurred and noisy data can significantly improve the results of experiments potentially leading to new and exciting scientific discoveries. While the benefits of the mathematical methods that make this possible are relatively well known, robust and easy-to-use software tools that implement these techniques are extremely rare. The Compressed Sensing for Magnetic Resonance Imaging and Cosmology (COSMIC) project (\url{http://cosmic.cosmostat.org/}) was funded by the Fundamental Research Division (DRF) at the French Alternative Energies and Atomic Energy Commission (CEA) to provide precisely these tools. 

COSMIC is a collaboration between two CEA groups with signal processing expertise: NeuroSpin, specialists in Magnetic Resonance Imaging (MRI), and CosmoStat, specialists in astrophysical image analysis. There is significant overlap in these fields, especially for astrophysical radio imaging that, like MRI, collects data in Fourier space. The primary output of this collaboration has been the development of the Python Sparse data Analysis Package (PySAP).

PySAP is an open-source software package written in Python that provides highly optimised sparse image transforms and a library of modular optimisation tools for solving linear inverse problems. While PySAP has been designed with specific applications to the MRI and astrophysics domains in mind, the versatility of the software and the universality of the mathematical techniques mean that it can also be applied to a variety of other imaging domains such as microscopy, tomography and echography.

Compared to existing inverse problem solving packages, such as SPAMS \citep{mairal:09a, mairal:09b, mairal:10, jenatton:10} and SigPy \citep{ong:19}, PySAP offers efficient implementations of specialised multiscale transforms. In particular, PySAP provides undecimated wavelet transforms, which are well suited to astrophysical images, and 3D wavelet transforms, which are of particular interest for MR image processing (see Sec.~\ref{sec:sparse2d}). PySAP also includes several detailed applications of these tools to data to facilitate user understanding (see Sec.~\ref{sec:app}).

This paper is organised as follows. Section~\ref{sec:pysap} provides a detailed description of the structure and features of PySAP with particular focus on the image transforms and optimisation tools. Section~\ref{sec:app} demonstrates practical applications of PySAP on MRI and astrophysical data. Finally, conclusions and plans for the future development of the package are presented.


\section{PySAP Features}
\label{sec:pysap}

In essence, the base PySAP package serves as a front-end that comprises several specialised modules. PySAP provides a simplified framework in which to combine these modules as well as managing file IO, visualisation and exception handling.

The core modules that provide the PySAP features are:
\begin{itemize}
  \item Sparse2D: Sparse Image Transforms
  \item ModOpt: Modular Optimisation Tools
  \item Plug-ins
\end{itemize}

\begin{figure}
  \centering
  \includegraphics[width=\columnwidth]{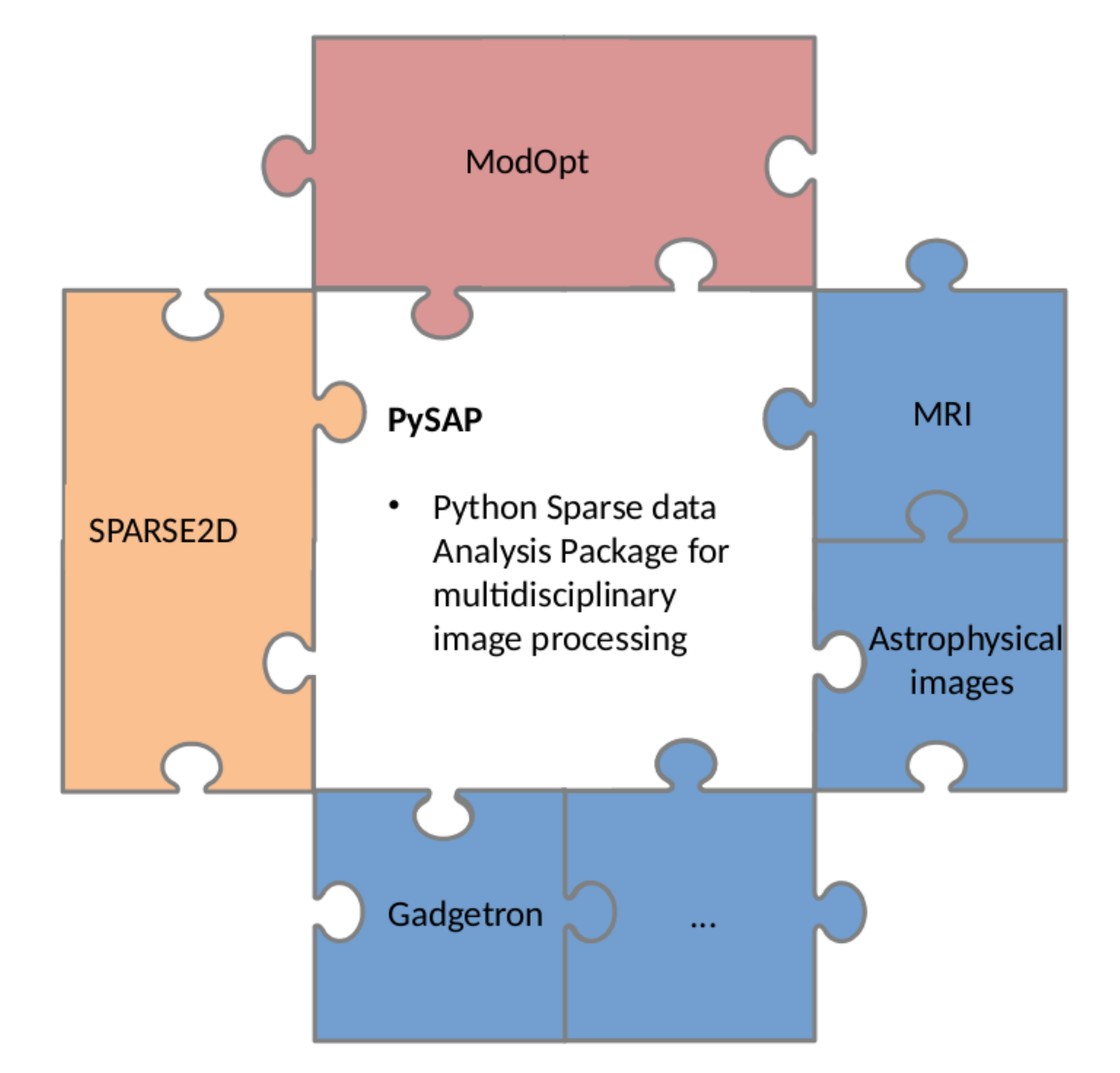}
  \caption{Illustration of the structure of the PySAP package. The SPARSE2D and ModOpt core libraries are represented in orange and red, respectively. The various plug-in applications appear in blue.}
  \label{fig:schema}
\end{figure}

\noindent Fig.~\ref{fig:schema} illustrates the core structure of the PySAP package. Each of these modules is described in detail in the following subsections.

\subsection{Sparse Image Transforms}
\label{sec:sparse2d}

The most essential tools for implementing compressed sensing or sparsity in signal processing problems are efficient dictionaries well suited to the data at hand. In practice, these dictionaries correspond to a series of data transforms ranging from wavelets to curvelets that can convert the data into a domain where the majority of the information is concentrated in very few non-zero coefficients, a concept called sparsity or compressibility.

Sparse2D is a C++ package, which forms part of iSAP (Interactive Sparse Astronomical data Analysis Packages), that provides a wide range of robust and efficient sparse transforms for 1D, 2D and 3D signals. In particular, the package includes a collection of undecimated wavelet transforms (UWT) that provide shift invariant properties for image reconstruction, such as the starlet transform \citep{starck:07} or the 7/9 UWT. These transforms are well documented in \citet{starck:15}. Sparse2D also includes 2D1D  and curvelets transforms, enabling the sparse decomposition of various types of data. A full list of the available transforms is provided in the \ref{sec:sparse2dtrans}. These software tools have been extensively tested on astrophysical data producing high quality results on a range of different topics \citep{bobin:14,leonard:14,ngole:15,lanusse:16}. The fact that this package relies on a set of fixed multiscale dictionaries means that the transforms are very computationally efficient and are therefore ideally suited to on-line MR image reconstruction \citep{ElGueddari:SPIE19}.

PySAP provides Python bindings for the Sparse2D C++ libraries, thus enabling fast and efficient implementation of the sparse transforms inside of a Python environment. This allows these tools to be more easily integrated into optimisation problems without any loss of performance (see section~\ref{sec:modopt}). Additionally, through the PySAP interface, Sparse2D transforms can be applied to MRI data separately on the real and imaginary parts. 

In addition to Sparse2D, PySAP also includes all of the transforms provided in PyWavelets \citep{lee:19}.

\subsection{Modular Optimisation Tools}
\label{sec:modopt}

Linear inverse problems, such as compressed sensing, are ill-posed because they are under-determined, \emph{i.e.} the number of measurements is far below the number of image pixels. To cope with this issue and make the inverse problem well posed, one usually resorts to regularisation. The image solution is then obtained as the minimiser of an optimisation problem. One of the main features of PySAP is a series of modular optimisation tools designed for solving linear inverse problems that comprise a subpackage called ModOpt. 

This package is particularly well suited for solving linear inverse imaging problems of the following form

\begin{equation}
	\mathbf{y} = \mathbf{H}\mathbf{x} + \mathbf{n}
\end{equation}

\noindent where $\mathbf{y}$ is the observed image obtained from the detector in question, $\mathbf{H}$ is a degradation matrix that could constitute blurring, sub-sampling, distortion, \emph{etc.}, $\mathbf{x}$ is the true image that one aims to recover and $\mathbf{n}$ is noise.

ModOpt provides robust and extremely flexible implementations of cutting-edge optimisation algorithms such as Forward--Backward, FISTA \citep{beck:09}, Generalised Forward--Backward \citep{raguet:11}, Condat--V\~u \citep{condat:13,vu:13} and POGM$^\prime$ \citep{kim:17}. For instance, these algorithms have been compared for MR image reconstruction in~\citep{ramzi:19}.
The flexibility of these implementations is provided via means of Python class composition. All of the proximity and linear operators as well as the gradient utilised by a given algorithm can be provided as class instances that inherit a parent structure to ensure smooth cohesion. The modularity of this approach means that any potential bug can be easily identified and fixed, thus ensuring a well maintained and robust framework. Additionally, this structure facilitates the future implementation of virtually any optimisation algorithm.

Predefined proximity operators are provided for implementing sparse, low-rank and structurally sparse regularisation~(\emph{i.e.}, $\ell_1$, nuclear and mixed norms, respectively) as well as a positivity constraint, which is commonly required in image analysis problems. Tools are included that allow the automatic setting of the regularisation parameters using the noise properties of the observed data. New proximity operator instances can easily be generated using the parent class. A list of the proximity operators currently available in ModOpt is provided in Table~\ref{tab:prox}. This structure includes a method that automatically calculates a given operator's contribution to the overall cost of the optimisation problem at hand.

\begin{table*}
  \caption{List of proximity operators currently available in ModOpt. The k-support norm generalises the group-LASSO $\ell_{2,1}$ norm~\citep{yuan:06} for imposing structured sparsity with overlapping groups, typically in the context of calibrationless MR image reconstruction. A particular case of Ordered Weighted L1~norm (OWL) that implements structured sparsity is OSCAR regularisation~\citep{bondell:08,ElGueddari:ISBI19}.\label{tab:prox}}
  \begin{tabular}{ll}
  	\hline
    {\bf Proximity Operator} & {\bf Application} \\
    \hline
    Positivity & Image analysis \\
    $\ell_1$ Minimisation & Sparse regularisation \\
    $\ell_*$ Minimisation & Low-rank regularisation \\
    $\ell_2$ Minimisation & Ridge regularisation \\
    $\alpha\ell_1 + \beta\ell_2$ Minimisation~($\alpha, \beta \geq 0$) & Elastic-net regularisation \\
    Ordered weighted $\ell_1$ norm~\citep{zeng:14}  & $\ell_\infty$-based Structured sparsity regularisation\\
    $k$-support norm~\citep{argyriou:12} & $\ell_2$-based Structured sparsity regularisation \\
    \hline
  \end{tabular}
\end{table*}

The linear operator parent class enables the use of any of the sparse image transforms described in section~\ref{sec:sparse2d}, in fact this framework is flexible enough to allow the implementation of virtually any custom transform. The structure of the this class also requires the definition of the adjoint process for a given transformation.

A standard gradient implementation of the form
\begin{equation}
  \nabla F(\mathbf{x}) = \mathbf{H}^T(\mathbf{H}\mathbf{x} - \mathbf{y})
\end{equation}

\noindent is included, where $F(\mathbf{x})$ is a convex function of the form $F(\mathbf{x}) = \frac12\|\mathbf{H}\mathbf{x}-\mathbf{y}\|_2^2$. The parent class structure ensures that the gradient required for a given inverse problem can be easily implemented. As with the proximity operators, the gradient's contribution to the total cost is built into the class structure.

A cost function class is also provided that automatically sums up the contributions from the proximity and gradient operators. This class has a built-in framework to test for convergence up to a given tolerance. 

Finally, a reweighting class is provided to counteract the bias introduced into a given solution owing to the use of soft-thresholding in sparse regularisation. At present, the method of \citet{candes:07} is included. 

The combination of these tools enables the user to very quickly prototype robust codes for tackling a variety of inverse imaging problems.

\subsection{Plug-ins}

PySAP also provides application specific plug-ins. In this module algorithms and operators from ModOpt can be combined with Sparse2D transforms to develop tools for a given application. The objective being to produce user-friendly functions, designed to solve well defined problems, that can be applied directly to data.

At present, this module contains plug-ins that demonstrate the applicability of PySAP to astrophysical and MRI data~(\texttt{pysap-astro} and \texttt{pysap-mri}, respectively). For example, the MRI plug-in \texttt{pysap-mri}~\citep{ElGueddari:20} adds the ability to deal with non-Cartesian data using non-uniform or non-equispaced FFT tools, while the astrophysics plug-in provides easy-to-use tools for denoising or deconvolving survey images. A specific plug-in, called \texttt{pysap-data} hosts the data sets used in the examples provided, while another one, called \texttt{pysap-tutorial}, contains materials for hands-on sessions. In the near future, two supplementary plug-ins will be released for other imaging techniques such as electron tomography~(\texttt{pysap-comset}) \citep{lin:20}, and electron microscopy~(\texttt{pysap-emicro}).

The plug-in framework has been designed to promote collaboration by providing a template for creating new plug-ins for virtually any imaging domain.


\section{Practical Applications}
\label{sec:app}

\subsection{Astrophysical Images}
\label{sec:astro}

One straightforward application of PySAP on astrophysical data is to the problem of galaxy image deconvolution. Astrophysical images obtained with optical telescopes are subject to a blurring caused by internal factors, such as imperfections in the optical system, and external factors, such as the atmosphere for ground based instruments. The sum of these aberrations is commonly referred to as the Point Spread Function (PSF).

\begin{figure*}
	\centering
	\includegraphics[width=0.49\textwidth]{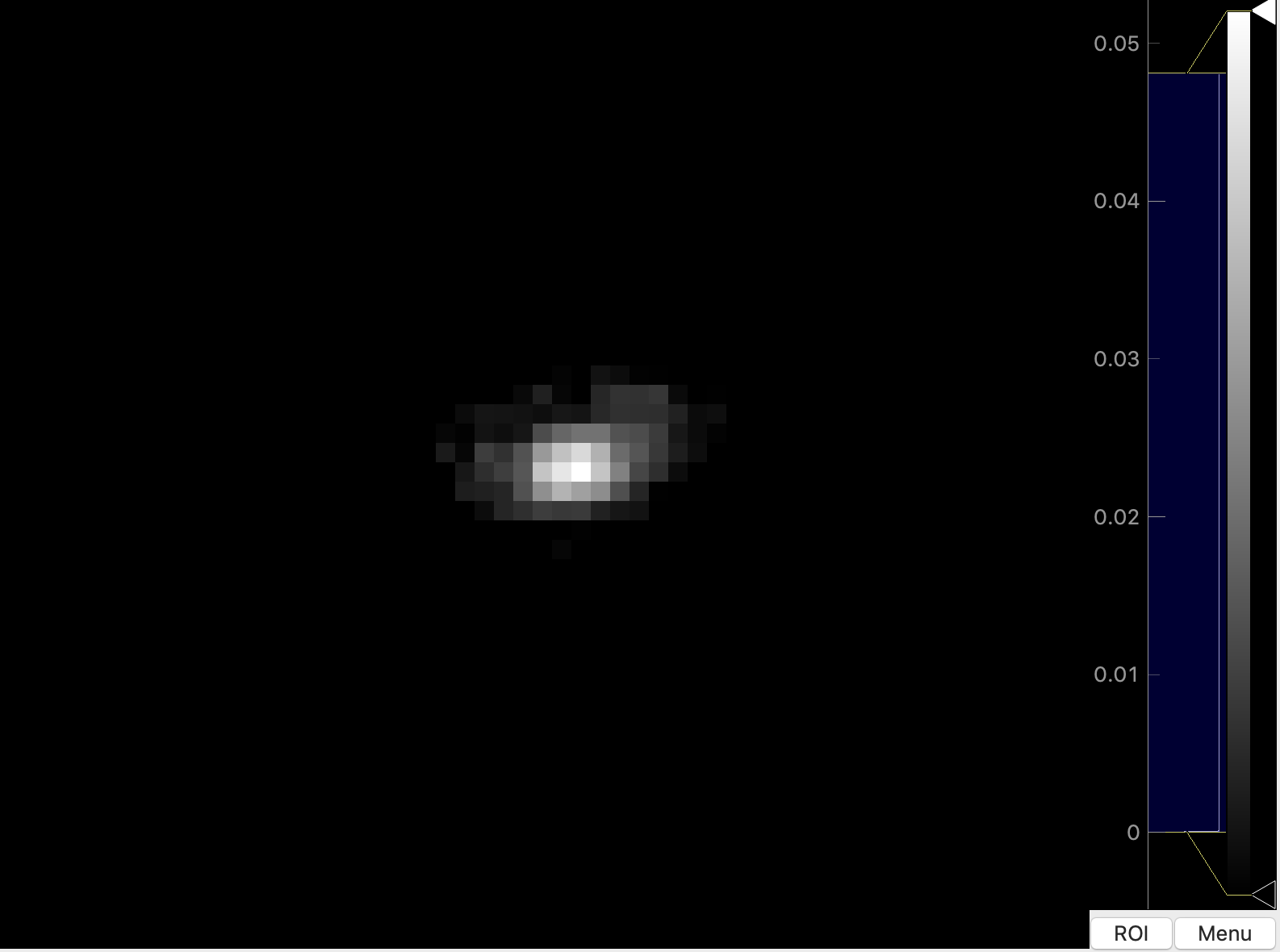}
    \includegraphics[width=0.49\textwidth]{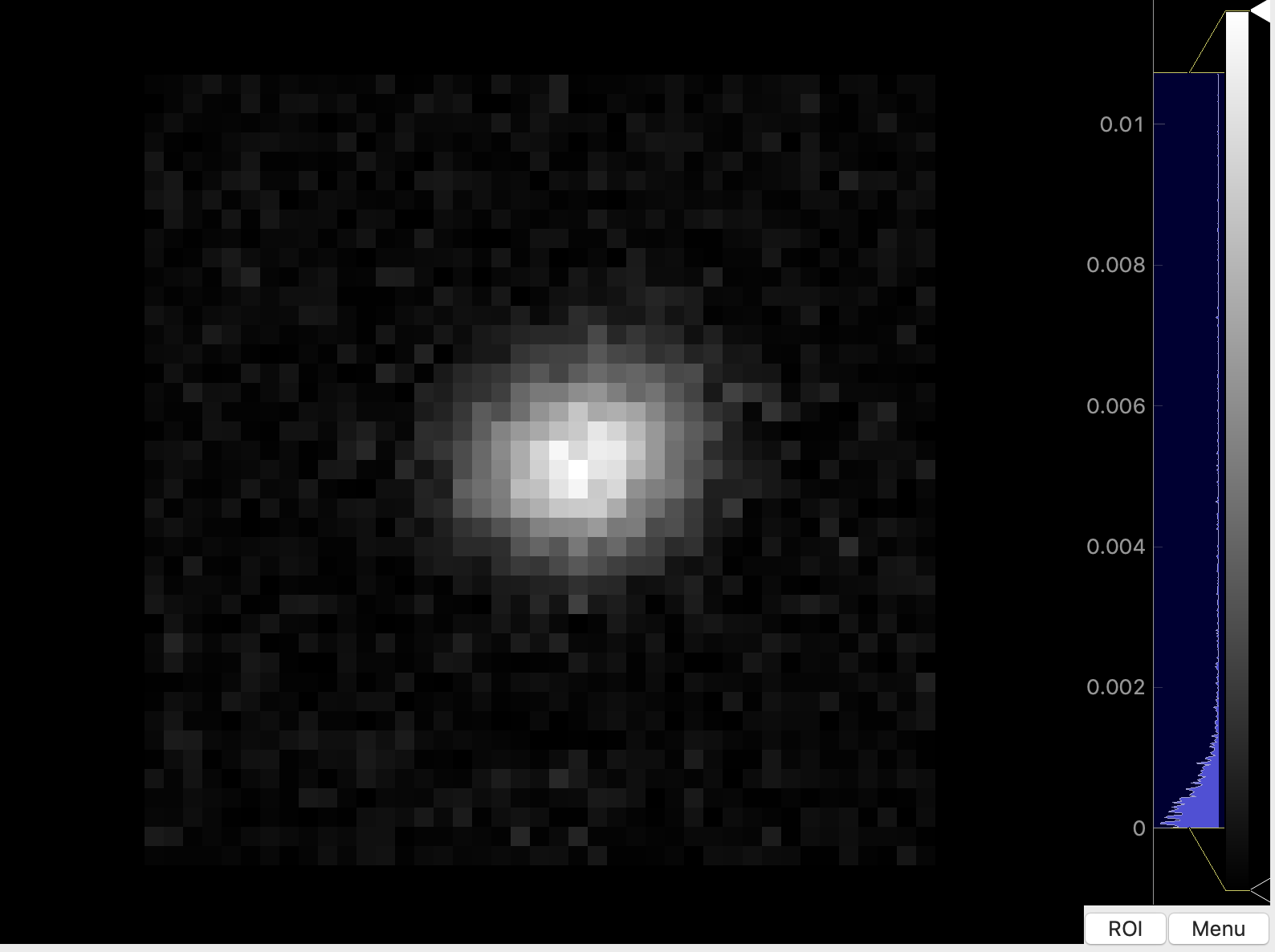}
    \includegraphics[width=0.49\textwidth]{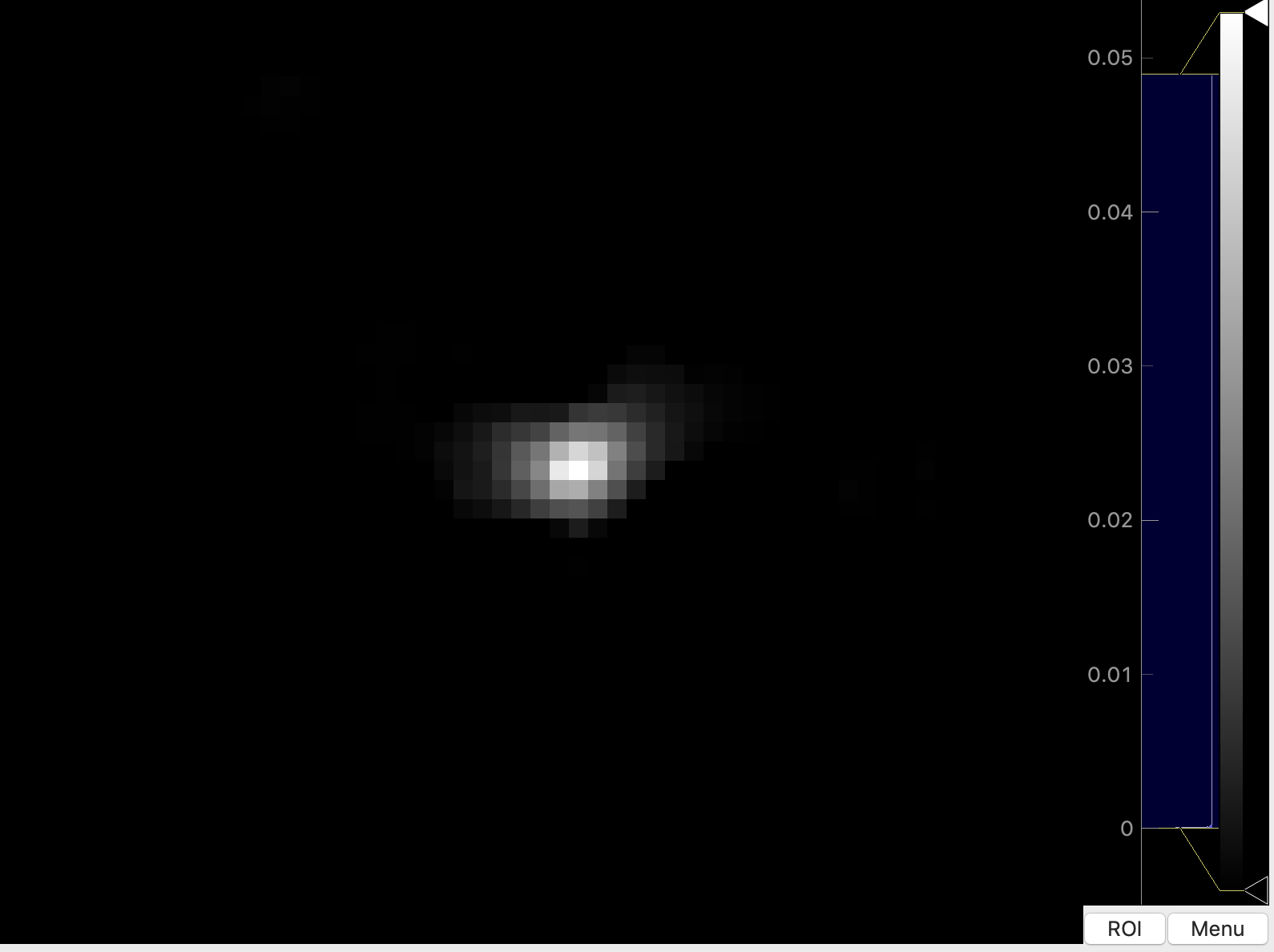}
    \includegraphics[width=0.49\textwidth]{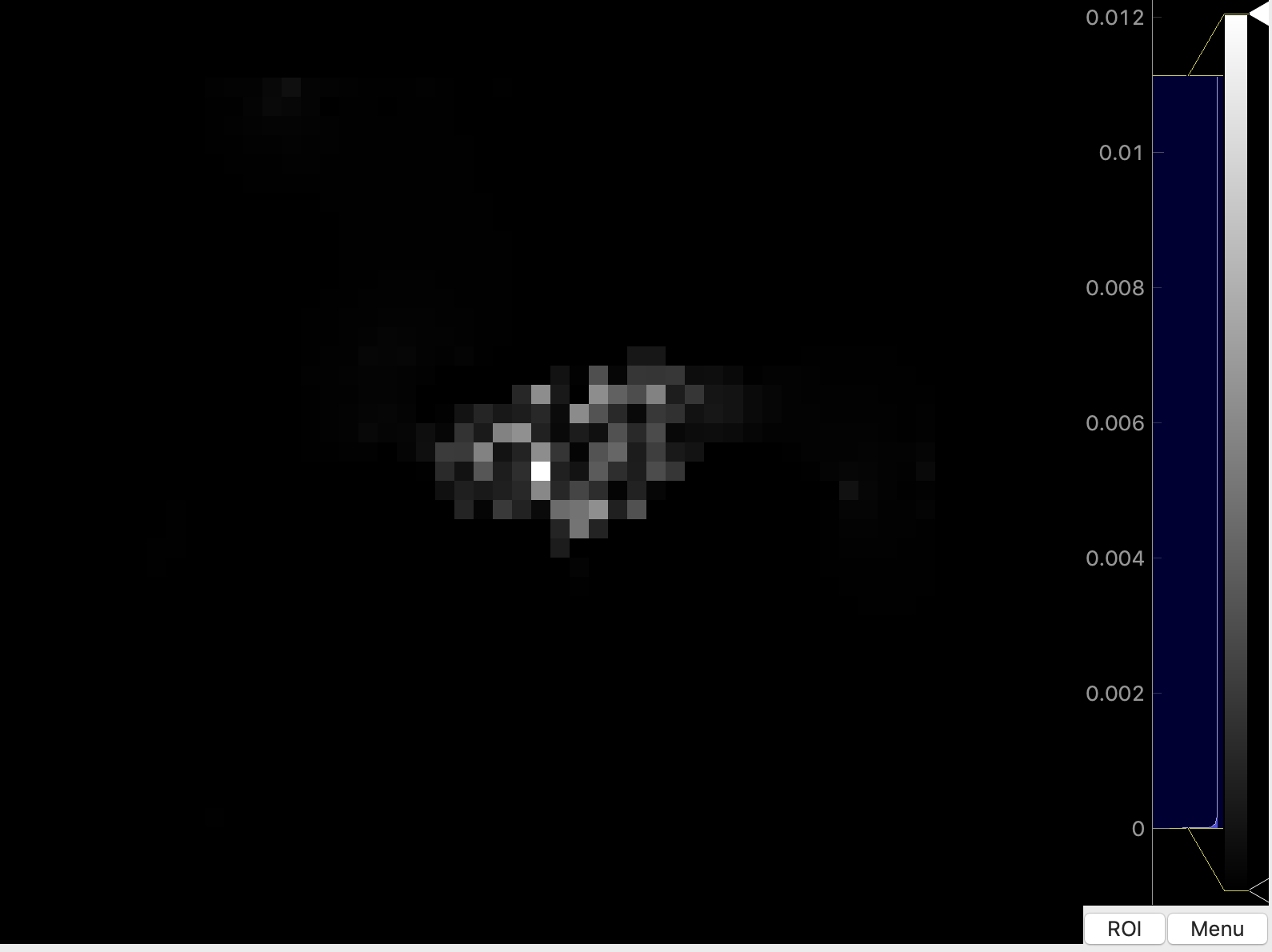}
    \caption{Example of galaxy image deconvolution using PySAP. \emph{Top left:} True galaxy image, \emph{Top right:} observed galaxy image, \emph{Bottom left:} deconvolved galaxy image, \emph{Bottom right:} deconvolution residual.}
    \label{fig:astro}
\end{figure*}

Removing the effects of the PSF from noisy observations amounts to solving a non-trivial inverse problem that requires the use of regularisation owing to the ill-conditioned nature of the degradation matrix, which corresponds to convolution with the PSF in this case. This problem can be solved using sparse regularisation following the same prescription described in \citet{farrens:17} using PySAP. A deconvolution example is provided in PySAP that demonstrates this process in a few lines of code. This example takes a COSMOS \citep{koekemoer:07, scoville:07a, scoville:07b} galaxy image that has been processed to remove noise \citep[see][]{farrens:17} as the true image that one aims to recover. An observation is then simulated by convolving this image with an anisotropic PSF and adding white Gaussian noise. This example performs deconvolution using the Condat--V\~u algorithm. An isotropic undecimated wavelet transform from Sparse2D is used for the linear operator, and a positivity constraint and soft-thresholding of the sparse coefficients are used as the proximity operators. The results of this example are shown in Figure~\ref{fig:astro}. 

\begin{figure*}
	\centering
	\includegraphics[width=0.49\textwidth]{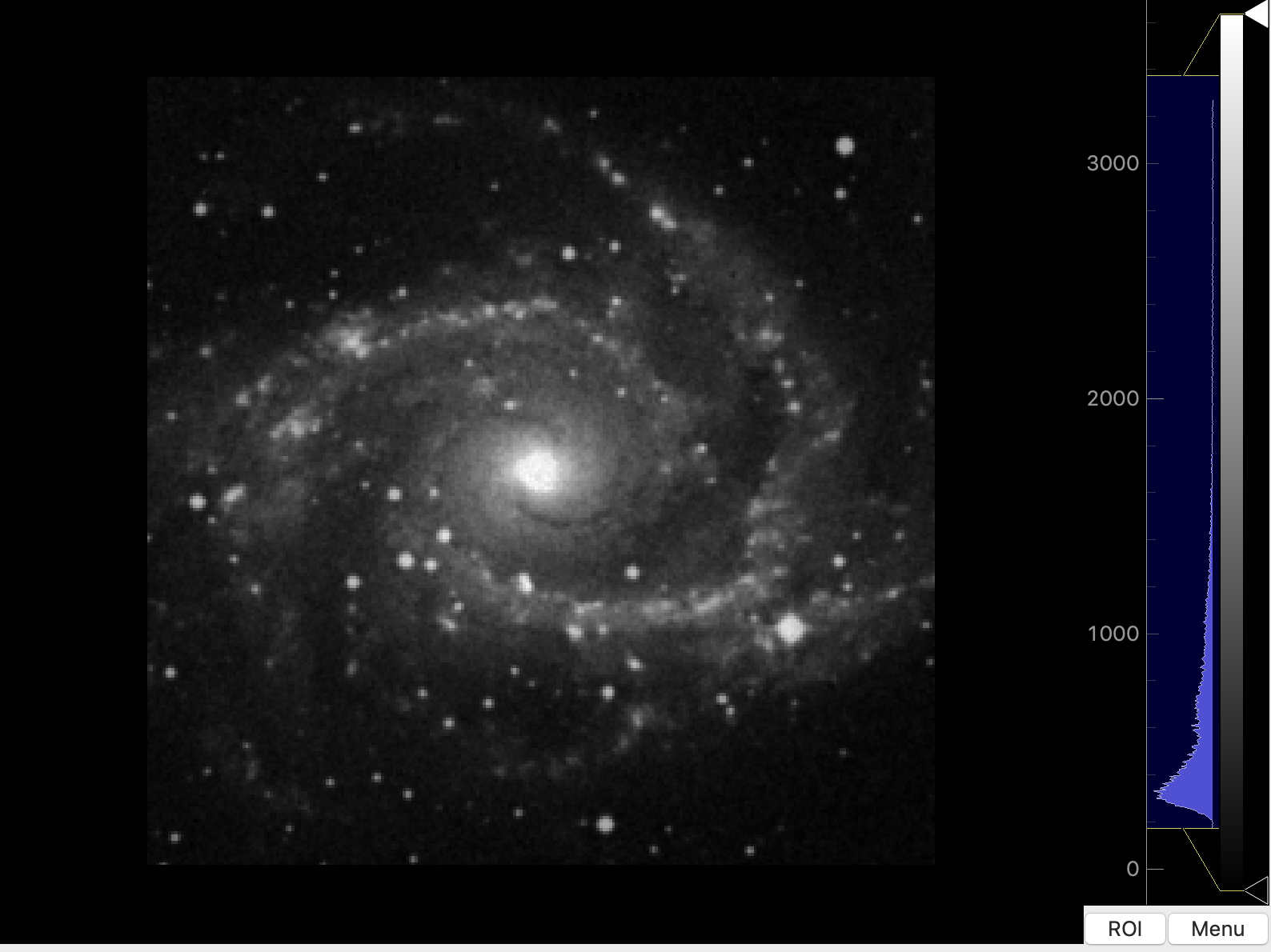}
    \includegraphics[width=0.49\textwidth]{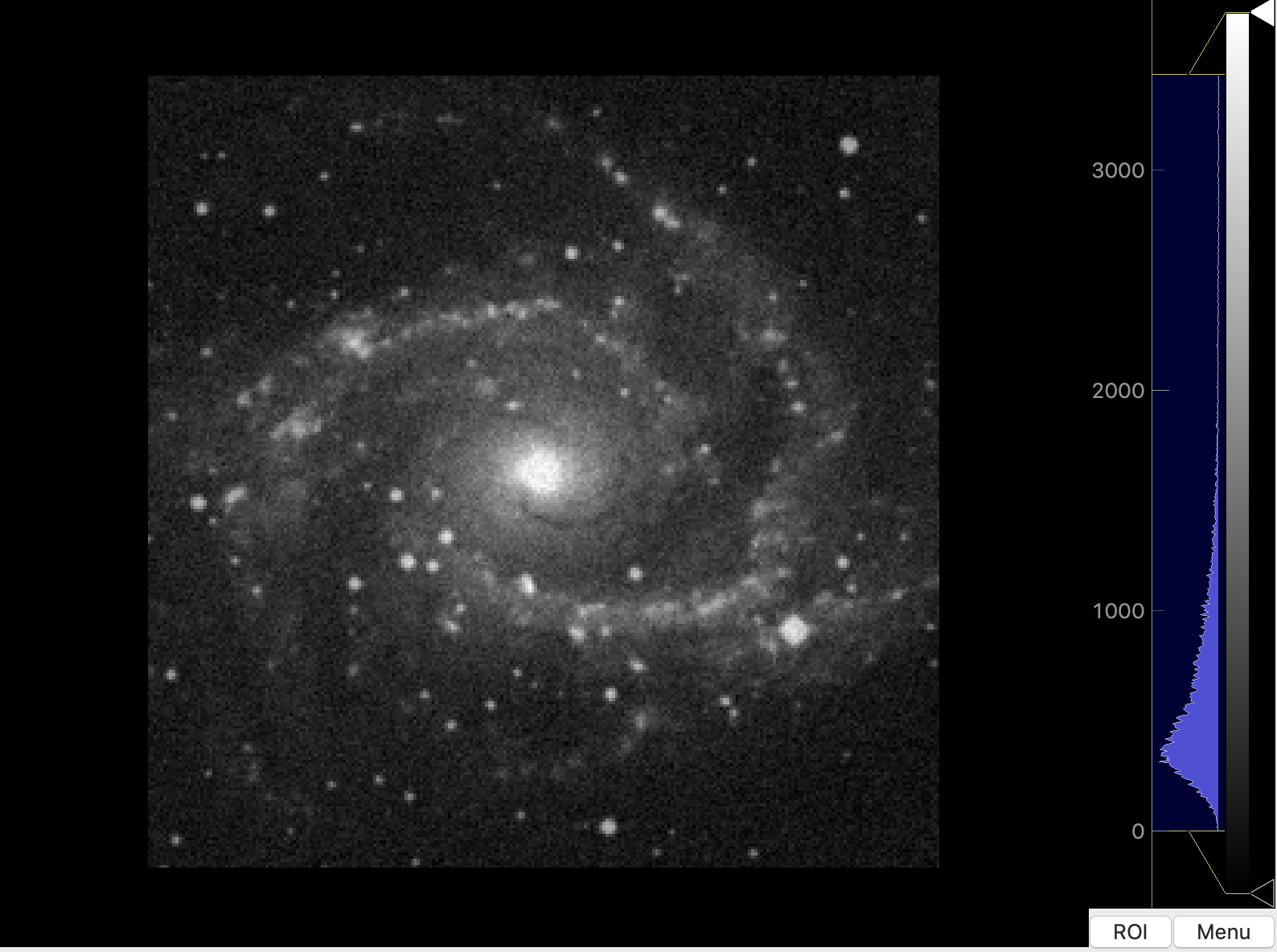}
    \includegraphics[width=0.49\textwidth]{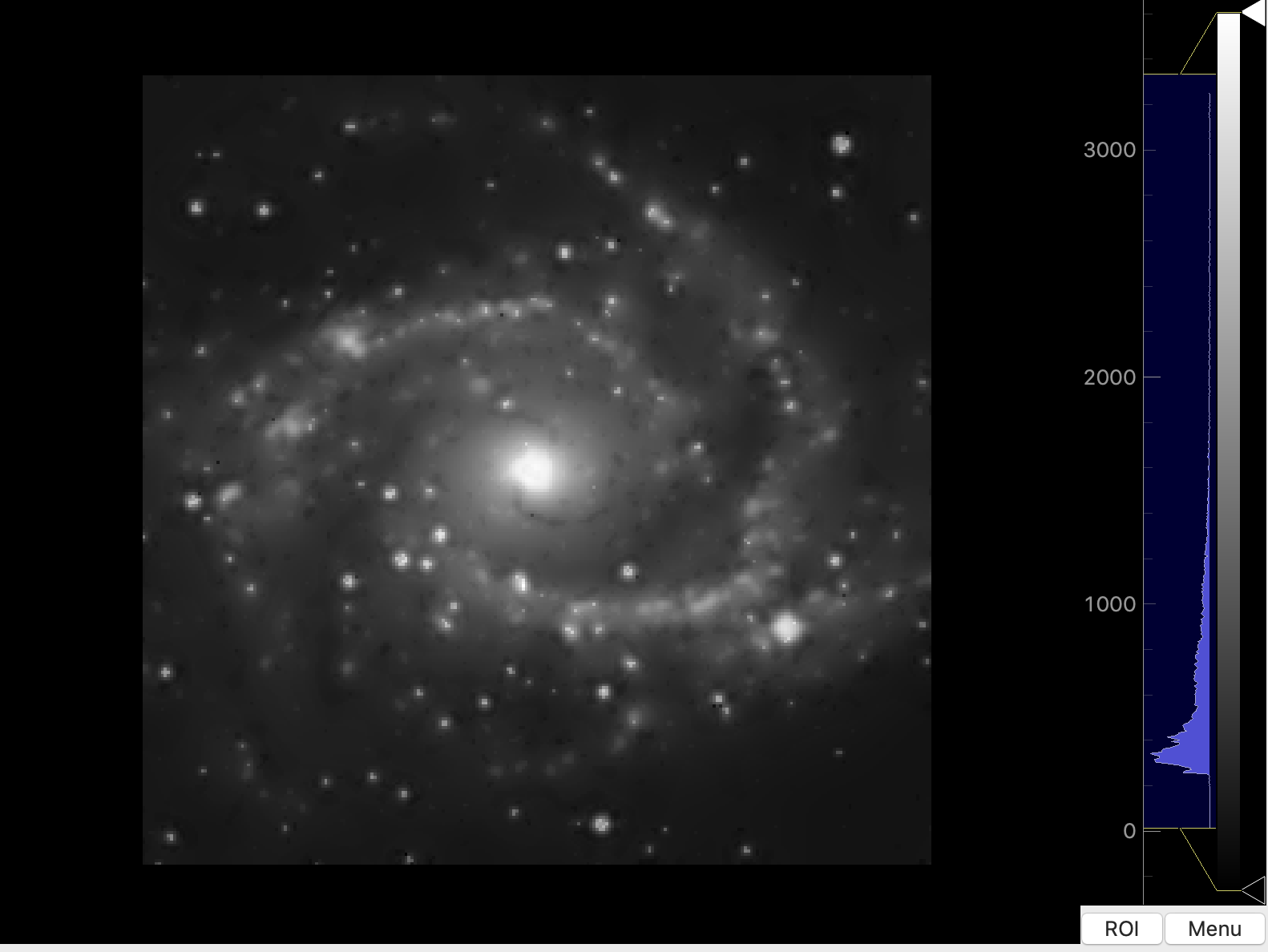}
    \includegraphics[width=0.49\textwidth]{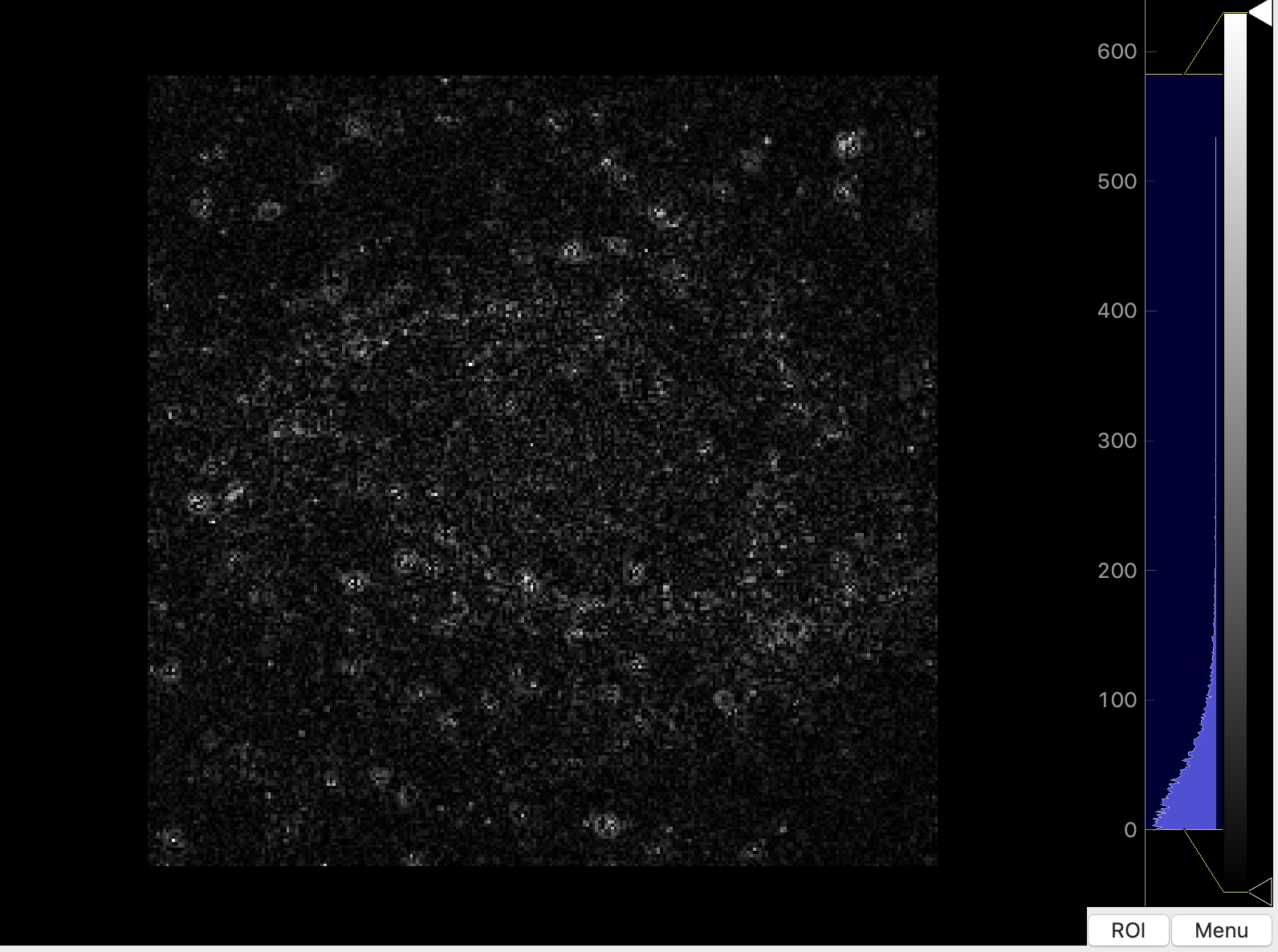}
    \caption{Example of galaxy image denoising using PySAP. \emph{Top left:} True galaxy image, \emph{Top right:} observed galaxy image, \emph{Bottom left:} denoised galaxy image, \emph{Bottom right:} denoising residual.}
    \label{fig:astro2}
\end{figure*}

Another application is simply removing noise from observations. This is a particularly challenging problem when the object in question contains important high-frequency spatial features that need to be preserved. Figure~\ref{fig:astro2} presents the results of denoising an image of the galaxy NGC2997 using PySAP. For this example white Gaussian noise is added to the clean image and then the same isotropic undecimated wavelet transform from Sparse2D is used to decompose the noisy image, which is in turn thresholded by weights learned from the noisy image itself.

Note that the data used for the examples presented in this paper are provided in PySAP~(in the {\tt pysap-data} plug-in). Therefore, all of the example outputs can be reproduced exactly by users.

\subsection{MRI}
\label{sec:mri}

MRI is probably one of the most successful applications of compressed sensing. The ability to reconstruct high-fidelity MR images from massively undersampled data in a short amount of time is of paramount importance. This is achievable as the data collected in the Fourier domain (called k-space in MRI) may be acquired using variable density sampling~(VDS) along a small number of trajectories~(or shots), either Cartesian or not~(\emph{e.g.} radial~\citet{jackson:1992}, spiral~\citet{meyer:1992} or more custom like Sparkling~\citet{lazarus:17b, Lazarus:MRM19}). The images can then be reconstructed using state-of-the-art optimisation algorithms. The idea is to take advantage of these time-saving strategies, not only to increase spatial resolution in anatomical imaging, but also to reduce sensitivity to motion. Importantly, PySAP is able to deal with both single and multi-channel 2D and 3D k-space data using GPU versions of NFFT operators. 3D VDS may be particularly relevant to improve the spatio-temporal resolution in functional MRI.

\captionsetup[subfigure]{labelformat=empty}
\begin{figure*}
    \centering
    \includegraphics[width=0.49\textwidth]{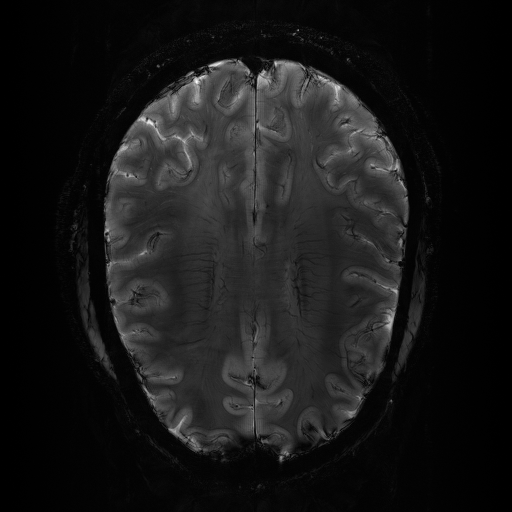}
    \includegraphics[width=0.49\textwidth]{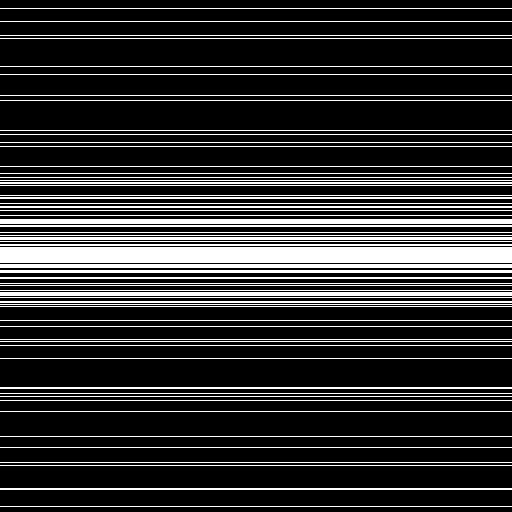}
    \includegraphics[width=0.49\textwidth]{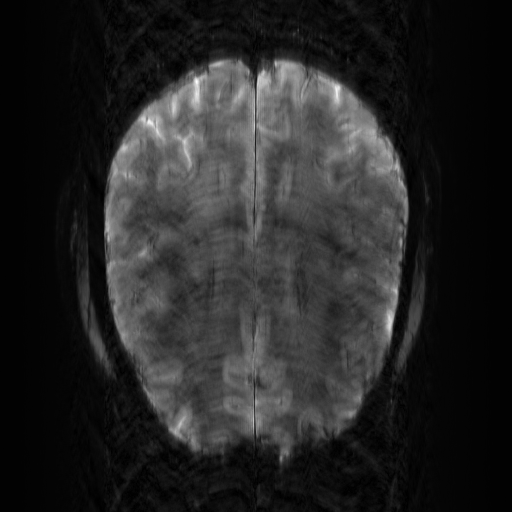}
    \includegraphics[width=0.49\textwidth]{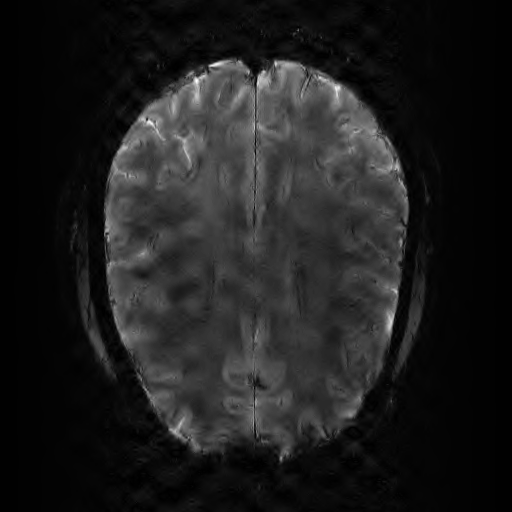}
    \caption{\label{fig:mri1}Cartesian MRI decimated wavelet-based reconstruction. \emph{Top left:} Cartesian reference, \emph{Top right:} K-space mask, \emph{Bottom left:} Zero-filled reconstruction ($SSIM=0.82$), \emph{Bottom right:} Decimated wavelet based reconstruction ($SSIM=0.91$).}
\end{figure*}

\begin{figure*}
    \centering
    \includegraphics[width=0.49\textwidth]{figures/MRI_Examples/MRI_Base_Image.png}
    \includegraphics[width=0.49\textwidth]{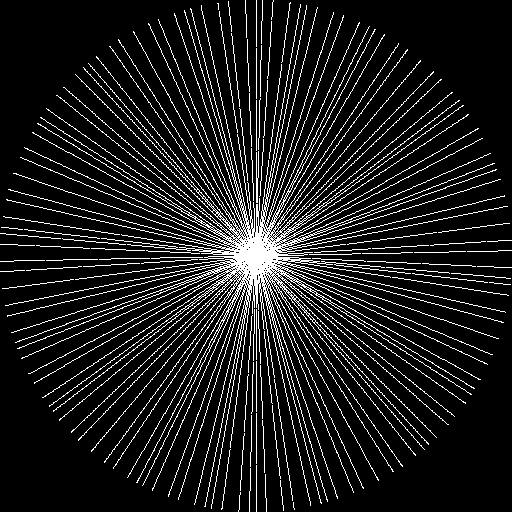}
    \includegraphics[width=0.49\textwidth]{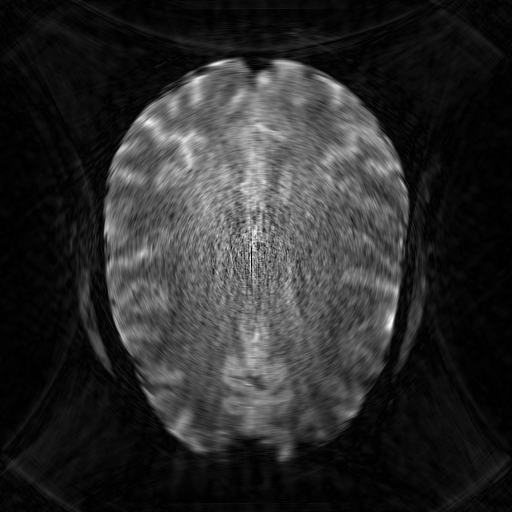}
    \includegraphics[width=0.49\textwidth]{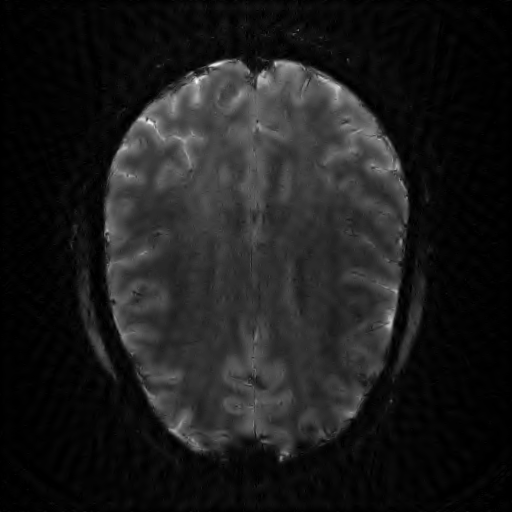}
    \caption{\label{fig:mri2}Non-cartesian MRI undecimated wavelet-based reconstruction. \emph{Top left:} Cartesian reference, \emph{Top right:} K-space mask, \emph{Bottom left:} Zero-filled reconstruction ($SSIM=0.67$), \emph{Bottom right:} Undecimated wavelet based reconstruction ($SSIM=0.92$).}
\end{figure*}

Take, for example, the problem of compressed sensing parallel imaging reconstruction. For this example let $L$ be the number of coils used to acquire the NMR signal, $n$ the image dimension and $N=n\times n$ be the number of pixels of a 2D complex-valued image $\mathbf{x}\in \mathbb{C}^N$ to be reconstructed and $M$ the number of samples collected per channel during acquisition. We denote by $\mathbf{y}_\ell\in \mathbb{C}^M$ 
the complex-valued data recorded by the $\ell^{\text{th}}$ channel, $\mathbf{S}_\ell\in \mathbb{C}^{N \times N}$ the corresponding diagonal sensitivity matrix. This matrix $\mathbf{S}_\ell$ reflects how well the $\ell$th receiver coil captures part of the object $\mathbf{x}$.
Let $F$ be the Fourier transform and $\Omega\subset\{1, \ldots, N\}$ the sampling pattern in the $k$-space, with $|\Omega|=M\ll N$. The CS-PI acquisition model thus reads:

\begin{align}\label{mri:pmri_model}
  \mathbf{y}_{\ell} &=  {\cal F}_\Omega\mathbf{S}_\ell\,\mathbf{x} + \mathbf{n}_\ell, \quad
  \forall \ell=1:L,
\end{align}

\noindent where $\mathbf{n}_\ell$ is additive zero-mean Gaussian noise of variance $\sigma^2_\ell$, which can be characterised by a separate scan~(without RF pulse) considering the same bandwidth as the prospective CS acquisition. In the case of Cartesian undersampling, ${\cal F}_\Omega = \mathbf{\Omega} \mathbf{F}$ where $\mathbf{\Omega}$ is a binary sampling mask with $M$ non-zero entries and $\mathbf{F}$ is the fast Fourier transform~(FFT). In the case of non-Cartesian undersampling, ${\cal F}_\Omega$ is the Non-equispaced Fourier transform~(NFFT). In the case where one assumes the sensitivity matrices $(\mathbf{S}_\ell)_{\ell=1}^L$ are known in advance, this problem can easily be solved using sparse regularisation following the same prescription described in \citet{chaari:11,guerquin:11} using PySAP in just a few lines of code. In the context of VDS, one may extract low frequency information from k-space data $(\mathbf{y}_\ell)_{\ell=1}^L$ to estimate the sensitivity maps $(\mathbf{S}_\ell)_{\ell=1}^L$ prior to reconstruction: this is called self-calibrated MR image reconstruction and has been implemented in~\citet{ElGueddari:SAM18}.

Several example applications to MR data are provided in the PySAP package. A first example is shown in Figure~\ref{fig:mri1}. It shows the reconstruction of an MR image from retrospectively undersampled k-space data. The original Cartesian data were collected in vivo on a healthy volunteer at 7~Tesla~(Magnetom Siemens scanner, Erlangen, Germany) using a 32-channel (Nova Medical Inc., Washington, MA, USA) coil~($L = 32$) and a 2D T2*-weighted GRE sequence~(see details in~\citep{Lazarus:MRM19}). To illustrate CS reconstruction algorithms, we actually used the reference image obtained as the square root of the sum of squares of 32 channels~(see Fig.~\ref{fig:mri1} top-left), which we retrospectively undersample. In that context, we emulated a single receiver coil to get rid of the estimation of sensitivity maps. Note however that specific algorithms are provided to extract these sensitivity matrices and perform image reconstruction in the dedicated plug-in for MRI as described in~\citep{ElGueddari:SAM18, ElGueddari:ISBI19}.

For educational purposes, we first used a Cartesian mask that implements variable density sampling along the phase encoding~(vertical) direction. We kept only $N_{c}=98$ phase encoding lines out of $n=512$, leading to an undersampling factor $R=N/M$ equal to the acceleration factor in time $\text{AF}=n/N_c=5.22$. The FISTA algorithm~\citet{beck:09} was used for optimisation purposes with sparsity promoted with the decimated symmlet 8 transform. The image reconstructed in Figure~\ref{fig:mri1} using this strategy outperforms the zero-filled inverse FFT by almost 0.1 in terms of structural similarity~(SSIM) score~(0.91 vs 0.82).

A second example is depicted in Figure~\ref{fig:mri2}, where retrospective radial undersampling was applied with only $N_c=64$ shots out of $n=512$ leading to a downsampling factor of $R=\text{AF}=8$. The sparsity was promoted using an anisotropic undecimated wavelet transform from Sparse2D. The image reconstructed using this strategy outperforms the zero-filled inverse NFFT by 0.25 in terms of SSIM score~(0.92 vs 0.67).

Note that as in these examples we only performed retrospective undersampling, $R=\text{AF}$, however, in prospective acceleration, one may gain in image quality using oversampling over each shot which leads to $R < \text{AF}$.
Some codes spinets can be found in the PySAP documentation gallery:~\\ \url{https://python-pysap.readthedocs.io/en/latest/auto_gallery/gallery.html}.

\subsection{Gadgetron}
\label{sec:gadgetron}


Gadgetron natively supports Python packages and therefore PySAP can easily be installed on any MRI scanner where the Gadgetron framework is in place.


\section{Conclusions}
\label{sec:conclusions}

In this paper we have presented the image processing package PySAP, its principal features and example applications to MR and astrophysical images. In particular, examples demonstrate how PySAP can be applied to image processing problems such as denoising, deconvolution and compressed sensing employing state-of-the-art reconstruction algorithms and wavelet transforms. In each case the plug-in framework provides easy-to-use tools for solving these problems for specific applications. 

The flexibility and modularity of this package permit a wide range of possible future developments. In particular, we aim to continue to add new and cutting-edge optimisation algorithms, reweighting methods, \emph{etc.} We additionally aim to add further features for handling 4D data and optimising the computation time by exploring GPU implementations. Another important aspect to which we plan to dedicate effort is to integrating machine and deep learning techniques into the existing architecture.

One of the most exciting uses of PySAP comes from the Gadgedtron implementation. The universality of this system and the growing community mean that PySAP can readily be used at MRI scanners around the world, potentially leading to some fascinating developments in the biomedical imaging domain.

Finally, we intend to seek out new applications of this software in a variety of different fields. In fact, work has already begun on developing a PySAP plug-in for electron tomography and electron microscopy.


\paragraph{Reproducible research} In the spirit of reproducible research PySAP is made publicly available and fully open source. Documentation and installation instructions are available on the PySAP website (\url{https://python-pysap.readthedocs.io/}). The authors kindly request that any academic publications that make use of PySAP cite this paper.  


\section*{Acknowledgments}

The authors wish to acknowledge the COSMIC project funded by the CEA DRF-Impulsion call in 2016, the Cross-Disciplinary Program on Numerical Simulation~(SILICOSMIC project in 2018) of CEA, the French Alternative Energies and Atomic Energy Commission. They are also grateful to
the UnivEarthS Labex program of Sorbonne Paris Cite (ANR-10-LABX-0023 and ANR-11- ´ IDEX-0005-02), the European Community through the grant DEDALE~(contract no. 665044) within the H2020 Framework Program of the European Commission, the Euclid Collaboration, the European Space Agency and the support of the Centre National d'Etudes Spatiales. The authors additionally acknowledge Hana\'e Carri\'e and Fabrice Poupon for contributing to the development of this package. 


\bibliography{pysap}


\appendix

\section{Sparse2D Transforms}
\label{sec:sparse2dtrans}

\subsection{1D Transforms}

\begin{enumerate}
    \item Linear wavelet transform: a trous algorithm
    \item B1spline wavelet transform: a trous algorithm
    \item B3spline wavelet transform: a trous algorithm
    \item Derivative of a b3spline: a trous algorithm
    \item Undecimated Haar wavelet transform: a trous algorithm
    \item Morphological median transform
    \item Undecimated (bi-) orthogonal wavelet transform
    \item Non orthogonal undecimated transform
    \item Modified positive B3spline wavelet transform: a trous algorithm
    \item Pyramidal b3spline wavelet transform
    \item Pyramidal median transform
    \item Morlet's wavelet transform
    \item Mexican hat wavelet transform
    \item French hat wavelet transform
    \item Gaussian Derivative wavelet transform
    \item (bi-) orthogonal wavelet transform
    \item (bi-) orthogonal transform via lifting scheme
    \item Wavelet packets
    \item Wavelet packets from lifting scheme
    \item Wavelet packets using the a-trous algorithm)
    \item Pyramidal linear wavelet transform
\end{enumerate}

\subsection{2D and 2D1D Transforms}

\begin{enumerate}
    \item linear wavelet transform: a trous algorithm
    \item bspline wavelet transform: a trous algorithm
    \item wavelet transform in Fourier space
    \item morphological median transform
    \item morphological minmax transform
    \item pyramidal linear wavelet transform
    \item pyramidal bspline wavelet transform
    \item pyramidal wavelet transform in Fourier space: algo 1 (diff. between two resolutions)
    \item Meyer's wavelets (compact support in Fourier space)
    \item pyramidal median transform (PMT)
    \item pyramidal laplacian
    \item morphological pyramidal minmax transform
    \item decomposition on scaling function
    \item Mallat's wavelet transform (7/9 filters)
    \item Feauveau's wavelet transform
    \item Feauveau's wavelet transform without undersampling
    \item Line Column Wavelet Transform (1D+1D)
    \item Haar's wavelet transform
    \item half-pyramidal transform
    \item mixed Half-pyramidal WT and Median method (WT-HPMT)
    \item undecimated diadic wavelet transform (two bands per scale)
    \item mixed WT and PMT method (WT-PMT)
    \item undecimated Haar transform: a trous algorithm (one band per scale)
    \item undecimated (bi-) orthogonal transform (three bands per scale)
    \item non orthogonal undecimated transform (three bands per scale)
    \item Isotropic and compact support wavelet in Fourier space
    \item pyramidal wavelet transform in Fourier space: algo 2 (diff. between the square of two resolutions)
    \item Fast Curvelet Transform
    \item Wavelet transform via lifting scheme
    \item 5/3 on line and 4/4 on column
    \item 4/4 on line and 5/3 on column
\end{enumerate}

\subsection{3D Transforms}

\begin{enumerate}
    \item (bi-) orthogonal transform
    \item (bi-) orthogonal transform via lifting scheme
    \item A trous wavelet transform
\end{enumerate}

\subsection{Curvelet Transforms}

\begin{enumerate}
    \item RectoPolar Ridgelet Transform using a standard bi-orthogonal WT
    \item RectoPolar Ridgelet Transform using a FFT based Pyramidal WT
    \item RectoPolar Ridgelet Transform using a Pyramidal WT in direct space
    \item Finite ridgelet transform
    \item Slant Stack Radon transform + FFT based pyramidal WT.
    \item Slant Stack Radon transformand + bi-orthogonal WT
    \item Slant Stack Radon transformand + pyramidal WT in direct space
    \item Slant Stack Radon transformand + Undecimated Starlet WT
\end{enumerate}


\end{document}